\documentclass[mathleft]{an}
\usepackage{times}
\usepackage{graphicx}
\usepackage{url}
\usepackage{bm}
\graphicspath{{./fig/}{./png/}}

\newcommand{\EQ}{\begin{equation}}
\newcommand{\EN}{\end{equation}}
\newcommand{\EQA}{\begin{eqnarray}}
\newcommand{\ENA}{\end{eqnarray}}
\newcommand{\eq}[1]{(\ref{#1})}

\newcommand{\Eq}[1]{Eq.~(\ref{#1})}

\newcommand{\Sec}[1]{Sect.~\ref{#1}}

\newcommand{\Fig}[1]{Fig.~\ref{#1}}

\newcommand{\bra}[1]{\langle #1\rangle}

\newcommand{\meanUU}{\overline{\vec{U}}}

\newcommand{\eee}{\hat{\bm{e}}}
\newcommand{\xxx}{\hat{\bm{x}}}
\newcommand{\yyy}{\hat{\bm{y}}}
\newcommand{\UU}{{\bm{U}}}
\newcommand{\uu}{{\bm{u}}}

\newcommand{\ff}{{\bm{f}}}

\newcommand{\kk}{{\bm{k}}}

\newcommand{\xx}{{\bm{x}}}

\newcommand{\nab}{\mbox{\boldmath $\nabla$} {}}
\newcommand{\OO}{\mbox{\boldmath $\Omega$} {}}

\newcommand{\SSSS}{\mbox{\boldmath ${\sf S}$} {}}

\newcommand{\ii}{{\rm i}}

\newcommand{\DD}{{\rm D} {}}

\newcommand{\const}{{\rm const}  {}}

\def\Rey{\mbox{\rm Re}}

\def\Co{\mbox{\rm Co}}
\def\cs{c_{\rm s}}
\def\kf{k_{\rm f}}
\def\kone{k_{\rm 1}}
\def\urms{u_{\rm rms}}

\def\half{{\textstyle{1\over2}}}

\def\onethird{{\textstyle{1\over3}}}

\newcommand{\yapj}[3]{: #1, {ApJ} {#2}, #3}

\newcommand{\yan}[3]{: #1, {AN} {#2}, #3}

\newcommand{\yana}[3]{: #1, {A\&A} {#2}, #3}

\newcommand{\ygafd}[3]{: #1, {GApFD} {#2}, #3}
\newcommand{\yjfm}[3]{: #1, {JFM} {#2}, #3}
\newcommand{\ypf}[3]{: #1, {PhFl} {#2}, #3}

\newcommand{\yprl}[3]{: #1, {Phys Rev Lett} {#2}, #3}

\newcommand{\ymn}[3]{: #1, {MNRAS} {#2}, #3}

\newcommand{\yjour}[4]{: #1, {#2} {#3}, #4}

\newcommand{\ybook}[3]{: #1, {\it #2} (#3)}

\Pagespan{725}{}
\Yearpublication{2011}
\Yearsubmission{2010}
\Month{1}
\Volume{332}
\Issue{1}
\DOI{10.1002/asna.200811027}

\begin{document}

\title{Verification of Reynolds stress parameterizations from simulations}
\authorrunning{J. Snellman et al.}
\author{
J. E. Snellman\thanks{Corresponding author: Jan.Snellman@helsinki.fi}$^{1,2}$,
A. Brandenburg$^{2,3}$,
P. J. K\"apyl\"a$^{1,2}$,
M. J. Mantere$^{1}$,
}
\institute{
$^1$Department of Physics, Gustaf H\"allstr\"omin katu 2a (PO Box 64),
    FI-00014 University of Helsinki, Finland\\
$^2$NORDITA, AlbaNova University Center, Roslagstullsbacken 23,
    SE-10691 Stockholm, Sweden\\
$^3$Department of Astronomy, Stockholm University,
    SE-10691 Stockholm, Sweden
}

\received{2010 Oct 22}  \accepted{2010 Nov 18}
\publonline{2010 Dec 30}

\abstract{%
We determine the timescales associated with turbulent diffusion and
isotropization in closure models using anisotropically forced and
freely decaying turbulence simulations and to study the
applicability of these models.  We compare the results from
anisotropically forced three-dimensional numerical simulations with
the predictions of the closure models and obtain the turbulent
timescales mentioned above as functions of the Reynolds number.  In
a second set of simulations, turning the forcing off enables us to
study the validity of the closures in freely decaying turbulence.
Both types of experiments suggest that the timescale of turbulent
diffusion converges to a constant value at higher Reynolds numbers.
Furthermore, the relative importance of isotropization is found to be about 2.5
times larger at higher Reynolds numbers than in the more viscous regime.
}

\maketitle

\section{Introduction}
\label{Introduction}

The dynamics of many astrophysical large-scale flows such as solar and
stellar differential rotation are strongly controlled by velocity
correlations at smaller scales.  These correlations are referred to as
components of the Reynolds stress tensor.  It is well known that in
rotating stratified convection the Reynolds stress tensor is
anisotropic (Kippenhahn 1963), which then leads to the generation of
differential rotation (R\"udiger et al.\ 1980, 1989).  The Reynolds
stress is defined as the average of products of components of velocity
fluctuations, i.e., $R_{ij}=\overline{u_iu_j}$, where
$\uu=\UU-\meanUU$ is the fluctuation of the velocity $\UU$ about its
mean $\meanUU$.  Here and in the following, overbars denote mean
quantities, and for the purpose of this paper we shall restrict
ourselves to volume averages.

Of particular interest are the equations governing the evolution of
$R_{ij}$.  In the astrophysical context, such model equations have
been derived by Ogilvie (2003) and Garaud \& Ogilvie (2005); see also
K\"apyl\"a \& Brandenburg (2008), Snellman et al.\ (2009), and Garaud
et al.\ (2010).  Such equations contain all the linear effects such as
shear and rotation exactly.  They usually also contain a driving term,
$F_{ij}$, through which energy is injected into the system, as well as
viscous and turbulent damping terms.  Finally, there often is a term
that describes, in a somewhat more ad-hoc fashion, the return to
isotropy (Rotta 1951).  The latter is important if the off-diagonal
components happen to be different from zero due to some statistical
perturbation.  At least at the level of a thought experiment, one
might ask how the system returns to isotropy after the effects that
produced the anisotropy, e.g., rotation and stratification via the
$\Lambda$-effect, have been turned off.  Mathematically, the turbulent
damping corresponds to terms involving triple correlations of the
velocity while the term describing the return to isotropy comes from
the interaction between components of velocity and those of gradients
of the pressure with the velocity (Canuto 2009).  Thus, in the absence
of large-scale shear flows, rotation, gravity, or magnetic fields, we
have
\begin{equation}
\dot{R}_{ij}=F_{ij}-\tau^{-1}R_{ij}
-\tau_{\rm iso}^{-1}\left(R_{ij}-\onethird\delta_{ij}R\right),
\label{reynolds}
\end{equation}
where the dot denotes a time derivative, $R=R_{ii}$ is the trace of
$R_{ij}$, while $\tau$ and $\tau_{\rm iso}$ are the relevant time
scales describing turbulent diffusion and the return to isotropy.

Two very similar ways to characterizing these timescales have been
proposed, both of which assume proportionality to the eddy turnover
time, and $\kf$ is the wavenumber of the energy-carrying eddies.
$\tau_{\rm 0}=(\urms\kf)^{-1}$, where $\urms$ is the rms velocity.  In
the standard minimal $\tau$-approximation (hereafter MTA) (Blackman \&
Field, 2002, 2003) the return to isotropy is not accounted for, and
$\tau$ is assumed constant in time.  The value of $\tau$ can be
expressed in terms of $\tau_{\rm 0}$ by defining a Strouhal number,
${\rm St}$, via
\begin{equation}
\tau= {\rm St}\,\tau_{\rm 0}.
\label{mta}
\end{equation}
If the isotropization term is included in MTA, $\tau_{\rm iso}$ is,
like $\tau$, also considered constant.  In an approach used by Ogilvie
(2003), the rms velocity is written as $\urms=R^{1/2}$, and
dimensionless fit parameters are introduced to quantify $\tau$ and
$\tau_{\rm iso}$:
\begin{equation}
\tau^{-1}=c_1\kf R^{1/2},\quad
\tau_{\rm iso}^{-1}=c_2\kf R^{1/2}.
\label{assumption}
\end{equation}
Besides the non-vanishing isotropization term, the main difference
between these models is the nature of the eddy turnover time: in MTA
it is usually constant, while in the Ogilvie approach it depends on
the local and instantaneous value of $R$.  The latter model can be
thought of as an extension of the former to the case where $\urms$
varies.

There seems to be some diversity regarding the recommended choice of
the coefficients $c_1$ and $c_2$.  For the ratio $c_1/c_2$, Garaud \&
Ogilvie (2005) found the value 0.67, while in the additional presence
of magnetic fields, Ogilvie (2003) found 0.87, and Liljestr\"om et
al.\ (2009) found 0.86.  The work mentioned above has attempted to
compute these coefficients as fit parameters in models where
additional effects such as shear, rotation, and gravity are present.
Such effects may however distort the results for $c_1$ and $c_2$,
which characterize effects that are present even without the
aforementioned processes.

A goal of this paper is to determine the two non-dimensional
coefficients $c_1$ and $c_2$ using direct numerical simulations (DNS).
We compute $c_1$ and $c_2$ here by imposing an anisotropic forcing
term such that certain off-diagonal terms of its correlation matrix
are non-vanishing. We use two independent methods to estimate the
parameters: firstly, by comparing the steady state values for $R$ and
$R_{ij}$ to the strength of the forcing, and secondly by observing the
behavior of the system once the forcing is turned off, that is freely
decaying turbulence. The predictions of the MTA and the Ogilvie
approach regarding the behavior of the system in the latter case
differ from one other, thus allowing us to assess the assumptions
behind the two closures.

\section{The model}
\label{sec:model}

We consider here a fully compressible gas with an isothermal equation
of state for which the pressure $p$ is proportional to the density $\rho$
with $p=\rho\cs^2$, where $\cs=\const$ is the isothermal sound speed.
The computational domain is assumed Cartesian $\xx=(x,y,z)$ with triply
periodic boundary conditions.
In some of our decay calculations, we start from a run where the
Coriolis force is included, which is characterized by the angular
velocity vector $\OO=(0,0,\Omega)$.
The equation of motion and the continuity equation can then be written as
\EQ
{\DD\UU\over\DD t}=-\cs^2\nab\ln\rho
-2\OO\times\UU+\ff+{1\over\rho}\nab\cdot(2\nu\rho\SSSS),
\EN
\EQ
{\DD\ln\rho\over\DD t}=-\nab\cdot\UU,
\EN
where $\DD/\DD t=\partial/\partial t+\UU\cdot\nab$ is the advective derivative,
${\sf S}_{ij}=\half(U_{i,j}+U_{j,i})-\onethird\delta_{ij}\nab\cdot\UU$
is the traceless rate of strain matrix, commas denote partial differentiation,
$t$ is the time, and $\nu$ is the kinematic viscosity.
The forcing term is an adaptation of a previously used (Brandenburg 2001) 
isotropic nonhelical
forcing expression, $\ff^{\rm iso}$, which is monochromatic with wavenumber
$\kk$, whose modulus lies in a narrow band around an average wavenumber $\kf$,
and the forcing is $\delta$-correlated in time such that $\kk_{\rm f}(t)$
changes abruptly from one time step to the next.
The isotropic forcing function is written as
$\ff=N\ff_{\kk}e^{\ii\kk(t)\cdot\xx}$, where $N$ is a normalization factor,
and $\ff_{\kk}=\eee\times\kk$ (with random unit vector $\eee$) to ensure
that the forcing is solenoidal.
Both $\eee$ and $\kk$ are random and non-parallel to each other.
Next, we introduce a finite $xy$ correlation by writing the forcing term as
\EQ
\ff=\ff^{\rm iso}+\sigma(\xxx f^{\rm iso}_y+\yyy f^{\rm iso}_x),
\EN
where $\xxx$ and $\yyy$ are unit vectors in the $x$ and $y$ directions,
respectively, and $\sigma$ is a non-dimensional parameter measuring
the degree of anisotropy.
Note that
\EQ
f_x f_y=(1+\sigma^2)f^{\rm iso}_x f^{\rm iso}_y
+\sigma[(f^{\rm iso}_x)^2+(f^{\rm iso}_y)^2],
\EN
and since $f^{\rm iso}_x f^{\rm iso}_y$ vanishes on the average,
$f_x f_y$ has a positive definite mean.
This then implies that in the Reynolds equations \eq{reynolds}
the forcing tensor
\EQ
F_{ij}=\rho(u_i f_j+u_j f_i)
\EN
is also anisotropic with $F_{xy}\neq0$ on the average.

To compute the effective timescales we consider steady state
conditions in which case \Eq{reynolds} implies
\EQ
\tau^{-1}=\bra{F}/\bra{R},
\label{forcingres1}
\EN
with $F=F_{ii}$ being the trace of $F_{ij}$, and
\EQ
\tau^{-1}+\tau_{\rm iso}^{-1}=\bra{F_{xy}}/\bra{R_{xy}},
\label{forcingres2}
\EN
where angle brackets now denote time averages.

A relevant control parameter is the Reynolds number,
defined as
\EQ
\Rey=\frac{\urms}{\nu\kf},
\EN
which is varied between 3 and 200.
In some of the decay calculations that are initialized with
rotation, we used a Coriolis number, $\Co=2\Omega/\urms\kf$ of order unity.
In all other cases we have $\Co=0$.

\section{Results}

We have produced three-dimensional DNS models with anisotropic forcing
varying both the Reynolds number and also the effective wavenumber of
the forcing, $\kf$. Firstly, we determine $\tau^{-1}$ and
$\tau^{-1}_{\rm iso}$ by comparing the steady state values for $R$ and
$R_{ij}$ to the strength of the forcing in Sect.~\ref{steady}.
In these experiments the numerical resolution is $256^3$ meshpoints.
Secondly, we determine the inverse relaxation time
scales from freely decaying turbulence in Sect.~\ref{decay}.
Here, the numerical resolution is $128^3$ meshpoints.

\subsection{Anisotropically forced turbulence}\label{steady}

\begin{figure}[t!]\begin{center}
\includegraphics[width=\columnwidth]{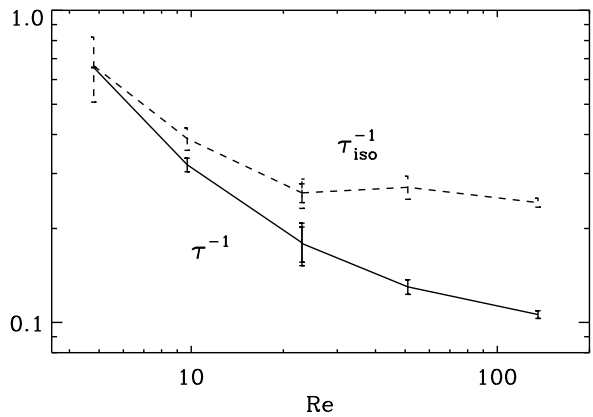}
\end{center}\caption[]{ 
Dependence of the inverse relaxation time scales (normalized by
the dynamical value $\tau_0^{-1}$ on $\Rey$.
Solid and dashed lines are for $\tau^{-1}$ and $\tau^{-1}_{\rm iso}$,
respectively.
}\label{psummary}\end{figure}

The inverse relaxation timescales $\tau^{-1}$ and $\tau^{-1}_{\rm iso}$
measured from anisotropically forced turbulence in a steady
state with varying Reynolds number and effective forcing wavenumber are
shown in \Fig{psummary}.
The results show a clear decline of $\tau^{-1}$ and $\tau^{-1}_{\rm iso}$
toward larger values $\Rey$.
At the same time,
$\tau^{-1}_{\rm  iso}$ is about 2.5 times larger than $\tau^{-1}$,
implying that $c_1/c_2\approx0.4$, which is somewhat smaller than the
values quoted in the literature; see \Sec{Introduction}.

\subsection{Decaying turbulence}\label{decay}

In this section we deterime the value of the timescales $\tau$ and 
$\tau_{\rm iso}$ and obtain another estimate for these parameters by studying 
freely decaying turbulence. 
We also compare the validity of the assumptions behind MTA and the Ogilvie
approach, since the closures predict decay behaviors that are different
in the two cases. By
letting the turbulence first achieve a saturated state and then turning off
the forcing in our DNS we get a time series that
can be compared with the predictions of the closure models. 
From Eq.~(\ref{reynolds}) we can easily derive the time evolution 
equation for the trace of the Reynolds tensor by summing over the 
diagonal components: 
\begin{equation}
\dot{R}=F-\tau^{-1} R,
\label{jrey}
\end{equation}
where the summation causes the contribution from the isotropization 
term to vanish. Let the forcing be set to zero at $t=t_0$ and let 
$R(t_0)=R^{(0)}$. 
If $\tau^{-1}$ is assumed constant in MTA, this approach predicts
exponential decay. By integrating Eq.~(\ref{jrey}) in this case we have
\begin{equation}
R=R^{(0)} e^{-(t-t_0)/\tau}.
\label{tmta}
\end{equation}
The Ogilvie approach, however, predicts inverse square-type decay:
\begin{equation}
R=\left[ \frac{1}{\sqrt{R^{(0)}}}+\frac{1}{2} c_1 k_f (t-t_0) \right]^{-2}.
\label{tog}
\end{equation}

\begin{table}
  \centering
  \caption[]{The model parameters $c_1$, $c_2$, $ \tau^{-1}$ and 
    $\tau_{\rm iso}^{-1}$ obtained from the DNS of freely decaying turbulence. The superscripts $b$ and 
    $l$ refer to the beginning and the late parts of the time series.}
  \vspace{-0.5cm}
  \label{tab:decayres}
  $$
  \begin{array}{p{0.06\linewidth}ccccccccrrr}
    \hline
    \noalign{\smallskip}
    Run & k_{\rm f}/\kone & \rm Re & \tau^{-1}/\tau_0^{-1} & c_1^b & c_1^l & \tau_{\rm iso}^{-1}/\tau_0^{-1} & c_2^b & c_2^l \\
    \noalign{\smallskip}
    \hline
    L1  & 3 & 53 & 0.11 & 0.12 & 0.16 & 0.08 & 0.14 & - \\
    L2  & 3 & 55 & 0.11 & 0.12 & 0.16 & - & - & -\\
    L3  & 3 & 61 & 0.12 &  0.14 & 0.16 & - & -  & - \\
    L4  & 3 & 79 & 0.12 &  0.14 & 0.17 & 0.04 & 0.07 & - \\
    L5   & 1.5 & 147 & 0.08 & 0.09 & 0.15 & - & - & - \\
    L6   & 10 & 14 & 0.19 & 0.23 & 0.27 & 0.06 & 0.08 & - \\
    L7   & 10 & 32 & 0.13 & 0.15 & 0.18 & - & - & - \\
    L8   & 3 & 113 & 0.11 & 0.13 & 0.16 & 0.08 & 0.08 & - \\
    L9   & 3& 191 & 0.10 &  0.11 & 0.15 & 0.06 & 0.10 & - \\
    \noalign{\smallskip}
    \hline
    F1   & 3 & 24 & 0.19 & 0.21 & 0.25 & 0.13 & 0.15 & - \\
    F2   & 3 & 53 & 0.13 & 0.13 & 0.19 & 0.19 & 0.23 & 0.05 \\
    F3   & 3 & 92 & 0.13 & 0.14 & 0.18 & 0.27 & 0.27 & 0.07 \\
    F4   & 1.5 & 55 & 0.13 & 0.15 & 0.25 & 0.32 & 0.32 & - \\
    F5   & 1.5 & 115 & 0.12 & 0.13 & 0.22 & 0.31 & 0.31 & 0.12 \\
    F6   & 1.5 & 192 & 0.12 & 0.13 & 0.20 & 0.10 & 0.11 & 0.04 \\
    F7   & 10 & 5 & 0.44 & 0.48 & 0.65 & 0.09 & 0.23 & - \\
    F8   & 10 & 13 & 0.23 & 0.27 & 0.32 & 0.13 & 0.15 & - \\
    F9   & 10 & 24 & 0.16 & 0.18 & 0.24 & 0.17 & 0.17 & - \\

   \end{array}
   $$ 
\end{table}

\begin{figure}[t!]\begin{center}
\includegraphics[width=\columnwidth]{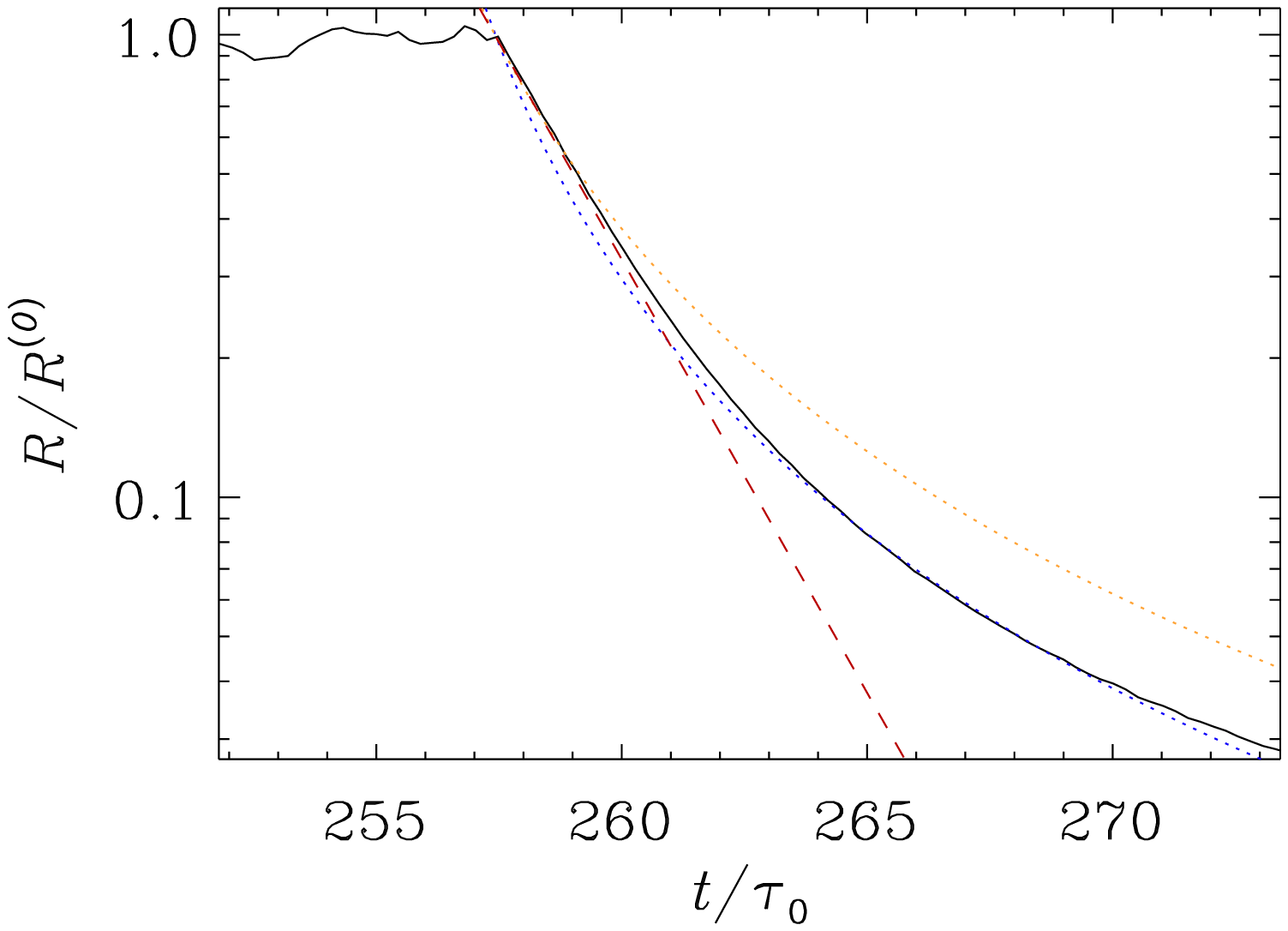}
\includegraphics[width=\columnwidth]{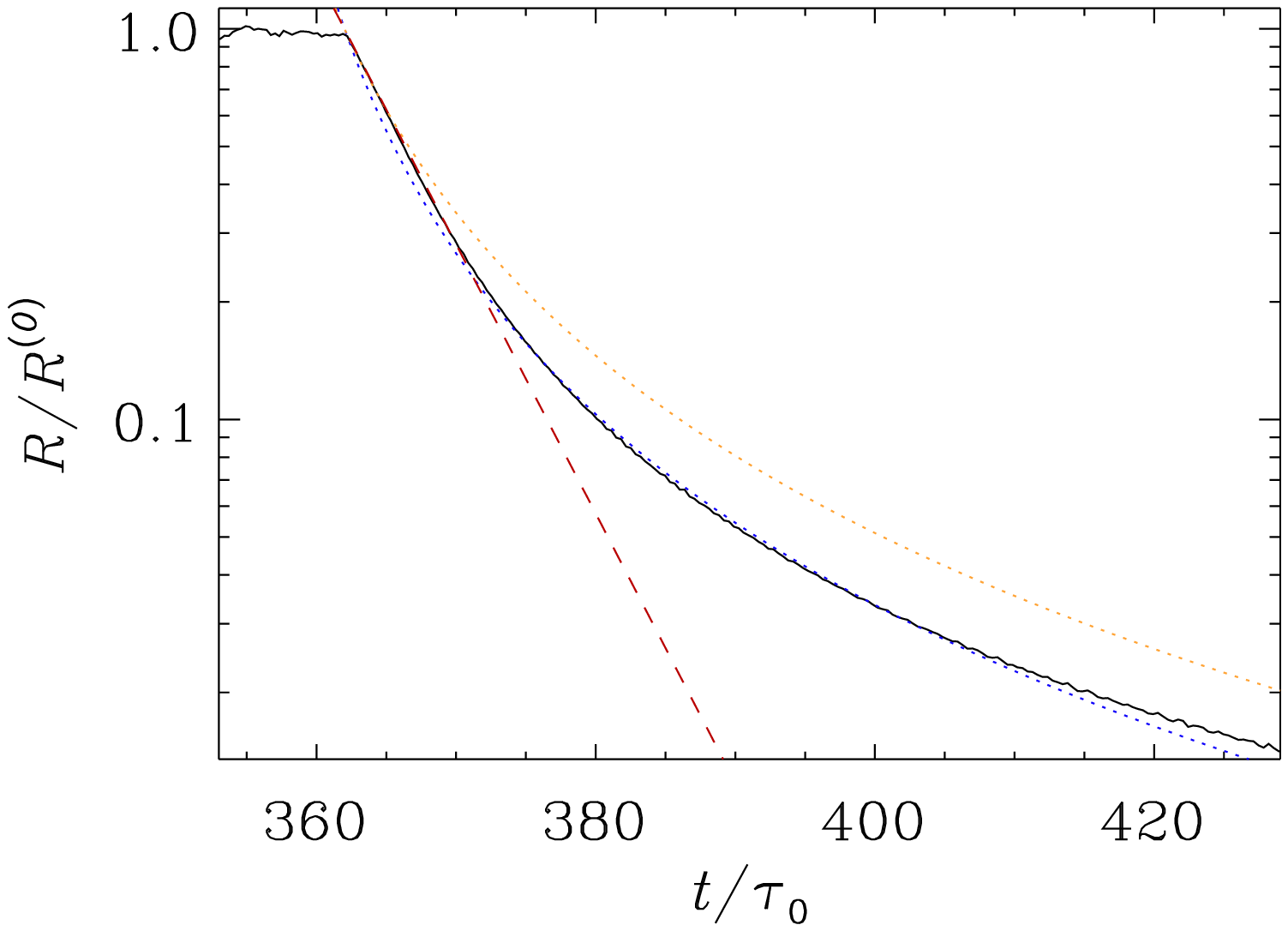}
\end{center}\caption[]{ 
The time evolution of $R$ in freely decaying turbulence. Dotted and dashed 
lines show the decay predictions of the 
Ogilvie model and MTA, respectively, with suitable values for $c_1$ and $\tau$
, and the solid line is the DNS time series. The upper panel shows the results 
from Run~F7 and lower panel results from Run~F9.
}\label{rel}\end{figure}

By plotting Eqs.~(\ref{tmta}) and (\ref{tog}) with the time series from
DNS the behavior of the closures can be
tested and the model parameters $c_1$ and $\tau$ estimated.
We have
performed two sets of runs, the results of which are summarized in
Table~\ref{tab:decayres}. In Set~F, we use the forcing
scheme described in Sect.~\ref{sec:model}, while the runs in Set~L were
made using anisotropic, nonhelical forcing in combination with rotation
($\Omega\neq0$ to) produce off-diagonal Reynolds stress components through
the $\Lambda$-effect; see K\"apyl\"a \& Brandenburg (2008) for a detailed 
description). The values listed in the table were obtained by fitting
Eqs.~(\ref{tmta}) and (\ref{tog}) to the DNS results. Two examples of
such a fit can be seen in Fig.~\ref{rel}. The solid lines represent
the DNS data, the dashed red lines the decay behavior predicted by the
MTA. The yellow and blue dotted lines are the corresponding prediction
of the Ogilvie closure with two different values for $c_1$, denoted
with $c^b_1$ and $c^l_1$ for the determination of which the beginning
and later parts of the DNS time series was used, respectively. The two
alternative fits for the latter model have been introduced because of
the changing nature of the process. As we can see, the decay generally
follows the exponential pattern at first, but in the later stages
power-law behavior similar to the prediction of the Ogilvie model takes place.
However, eventually the DNS results move away from both predictions.

This kind of changing behavior is observed in all of the decay models, 
and the temporal span of the validity of various predictions vary between 
the runs. This can be seen in Fig.~\ref{rel}, in which the upper
panel shows the fit to the DNS data from Run~F7, and the lower panel
shows a corresponding fit to the data from Run~F9:
while the exponential prediction of MTA seems to apply for approximately the 
same duration 
in both panels, the Ogilvie approach has clearly a different range of 
applicability.
Table~\ref{tab:decayres} lists the different fit parameters $c^b_1$, $c^l_1$
and $\tau^{-1}/\tau_0^{-1}$ obtained from the decay models. The
values for $c^b_1$ are generally very close to the values of
$\tau^{-1}/\tau_0^{-1}$, while $c^l_1$ tend to be somewhat larger.
Actually, if one puts $c_1=\tau^{-1}/\tau_0^{-1}$, the resulting curve has 
MTA prediction as a tangent at $t_0$.

Parameters $\tau_{\rm iso}$ and $c_2$ can be estimated by studying the decay of 
the off-diagonal components of the Reynolds stress. 
The time evolution equation for $R_{ij}$ in the forced non-diagonal case reads 
\begin{equation}
\dot{R}_{ij}=F_{ij}-(\tau^{-1}+ \tau_{\rm iso}^{-1})R_{ij}.
\label{jrxy}
\end{equation}
Now, let $R_{ij}(t_0)=R_{ij}^{(0)}$. Assuming $\tau_{\rm iso}$ constant in the 
case of MTA we have again exponential decay:
\begin{equation}
R_{ij}=R_{ij}^{(0)} e^{-(t-t_0)(\tau^{-1}+ \tau_{\rm iso}^{-1})}.
\label{tmtaxy}
\end{equation}
To get the corresponding result for the Ogilvie model one needs to use 
Eq.~(\ref{tog}) to solve for $\sqrt{R}$ and integrate over time. The final 
result reads
\begin{equation}
R_{ij}=R_{ij}^{(0)}\left[1+\frac{\sqrt{R^{(0)}}}{2} c_1 k_{\rm f} (t-t_0) \right]^{-2\frac{c_1+c_2}{c_1}}.
\label{togxy}
\end{equation}

\begin{figure}[t!]\begin{center}
\includegraphics[width=\columnwidth]{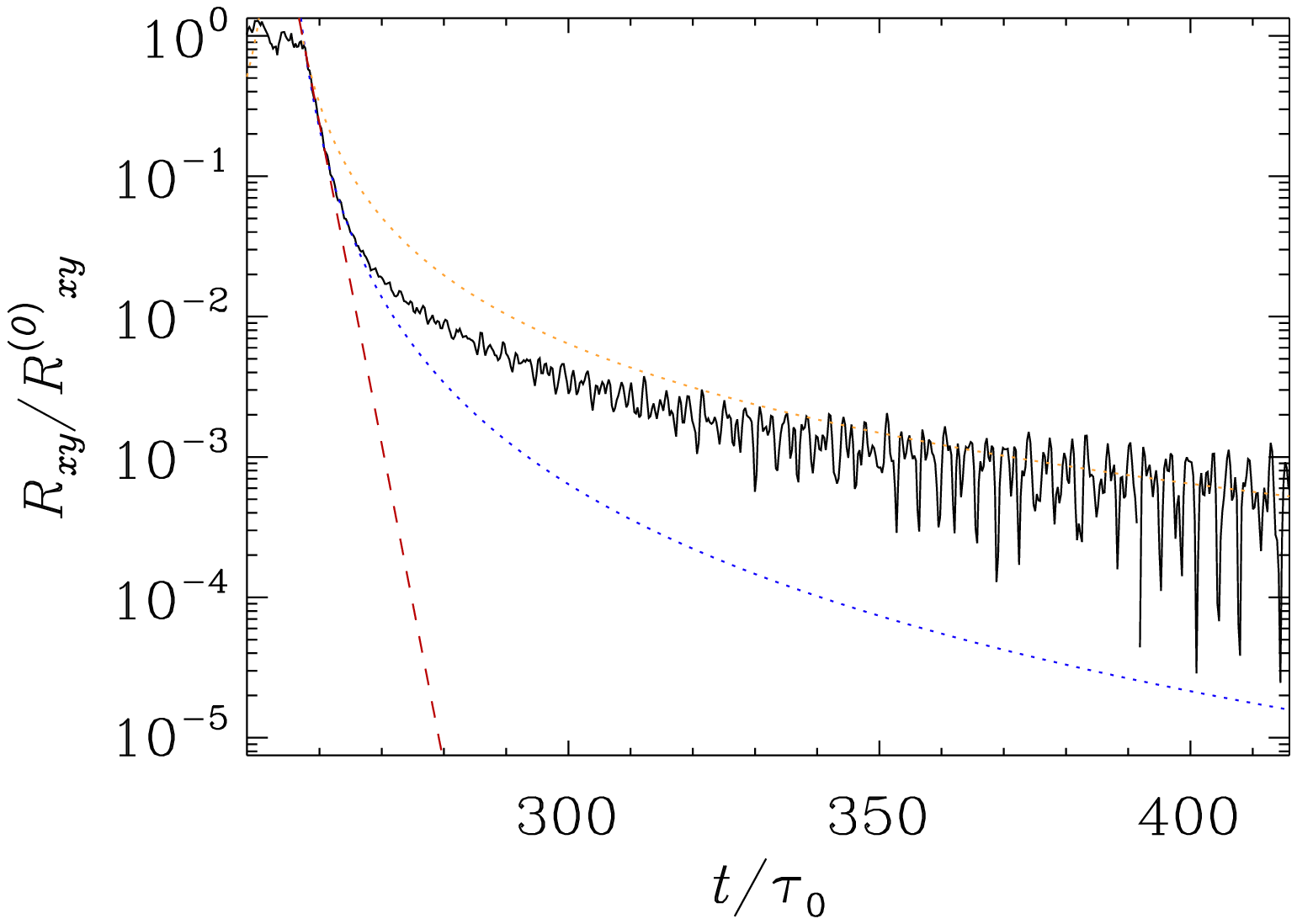}
\includegraphics[width=\columnwidth]{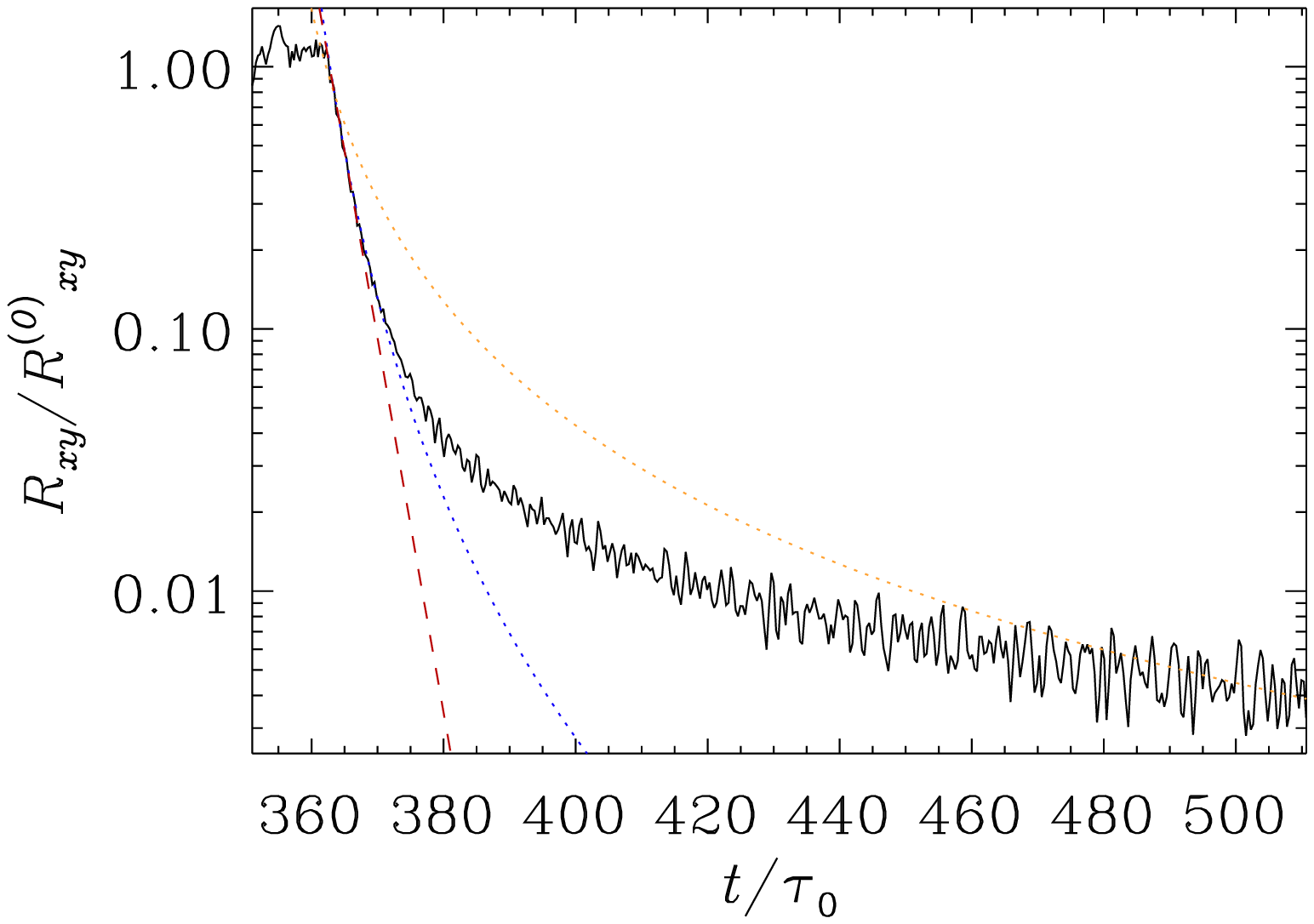}
\end{center}\caption[]{ 
The time evolution of $R_{xy}$ in freely decaying turbulence. 
Dotted and dashed lines show the decay prediction of the 
Ogilvie model and MTA, respectively, with suitable values for $c_2$ and $\tau$
, and the solid line is the DNS time series. The runs displayed are the same 
as in Fig.~\ref{rel}.
}\label{relxy}\end{figure}

The DNS results are compared with the predictions from the closure models in 
Fig.~\ref{relxy}. Again we show two alternative versions for the behavior of 
the Ogilvie model with different values for $c_2$, $c_2^b$ and $c_2^l$, with 
the same reasoning as with $c_1$. According to Eqs.~(\ref{tmtaxy}) and 
(\ref{togxy}), the decay of $R_{xy}$ depends on the relaxation parameters 
$\tau$ and $c_1$ as well as the dedicated isotropization parameters 
$\tau_{\rm iso}$ and $c_2$. Using the estimates for the relaxation 
terms obtained from the decay of $R$ we can determine the isotropization terms 
by treating them as the only free parameters of the models and finding a 
reasonable fit, like before. In the case of $c_2$ we have used the initial 
value $c_1^b$ for this purpose. 

The results for the isotropization terms are summarized in 
Table~\ref{tab:decayres}. 
A problem in many runs is that the fluctuations of the off-diagonal 
components of the Reynolds stresses can be larger than their average value, 
causing their sign to change frequently. In the decay phase the time series of 
these runs tend to contain strong oscillations right from the beginning.
The oscillations are similar to what can be seen in Fig.~\ref{relxy}, and 
they make finding an unambiguous fit very challenging.  In some cases a 
suitable fit would have required negative values for the parameter $c_2$. 
 For these cases, no value
is given in Table~\ref{tab:decayres}. This problem manifests itself mostly 
in Set~L. Thus, the most reliable results come
from Set~F, where $R_{xy}$ get non-zero mean values more
consistently, and fluctuations are not too large. 
We see that $\tau_{\rm iso}^{-1}/\tau_0^{-1}$ and
$c_2^b$ obtain very similar values, while $c_2^l$ is mostly very small or
zero. Equation~(\ref{togxy}) implies that with $c_2=0$ the decay of
the off-diagonal components of the Reynolds stresses should behave
like the decay of $R$ described by Eq.~(\ref{tog}), so the vanishing
of $c_2^l$ may indicate the isotropization switching off. But then
again, it is seen in Fig.~\ref{relxy} that even with vanishing $c_2$
the prediction becomes gradually worse as time progresses, and in the
lower panel the period of validity is restricted to a brief
intersection. Large fluctuations are another source of ambiguity near
the end of the time series.

\begin{figure}[t!]\begin{center}
\includegraphics[width=\columnwidth]{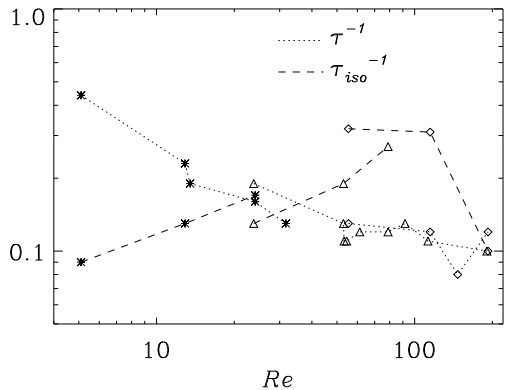}
\end{center}\caption[]{ 
Inverse relaxation time scales (normalized by
the dynamical value $\tau_0^{-1}$) as functions of $\Rey$ obtained 
from the decay models.
Dotted and dashed lines are for $\tau^{-1}$ and $\tau^{-1}_{\rm iso}$,
respectively. The diamonds represent runs with $\kf/\kone=1.5$,
triangles $\kf/\kone=3$ and asterisks $\kf/\kone=10$.
}\label{rec1}\end{figure}

Figure~\ref{rec1} contains the same results as Fig.~\ref{psummary},
but obtained for the decay models.
Due to the ambiguity of the results from the Set~L, only results from Set~F 
are shown for $\tau_{\rm iso}^{-1}$. 
In both figures the overall trend is similar: 
$\tau^{-1}$ is large with small Reynolds numbers, and decreases as $\Rey$ 
increases. Unlike in Fig.~\ref{psummary}, in Fig.~\ref{rec1} 
$\tau_{\rm iso}^{-1}$ generally increases with increasing $\Rey$, 
and eventually becomes greater than $\tau^{-1}$. It would seem that the 
results for $\tau^{-1}$ approach some constant value at high Reynolds 
numbers, but more simulations with higher Reynolds numbers would be needed 
to verify this.
Increasing $\tau^{-1}$ with decreasing $\Rey$ may explain, why the nature of 
the decline of $R$ changes in the decay models. If we take $\urms=R^{1/2}$ in 
the decay phase, the effective Reynolds numbers falls accordingly. This would 
mean that $\tau^{-1}$ changes during the simulation, leading to a different
behavior. 

\section{Conclusions}

In this study we have investigated anisotropically forced hydrodynamic
turbulence, and determined the timescales related to the diffusion and
isotropization processes from our DNS models. The obtained results
were compared to two different closure model predictions, namely the
minimal tau approximation and the Ogilvie approach.

Our results from the steady-state forced turbulence models show that
the values of $\tau^{-1}$, describing the diffusion process, and
$\tau^{-1}_{\rm iso}$, describing the isotropization process,
depend on $\Rey$ for small and intermediate values, but show clear
signs of convergence for larger values. In particular, it turns out
that $\tau^{-1}_{\rm iso}$ is clearly larger than $\tau^{-1}$, and
that their inverse ratio is around 0.4, which is somewhat less than
the results published earlier in the literature.

Our models of freely decaying turbulence show that, while the decay is
exponential at first, as predicted by the MTA with a constant $\tau$,
it deviates from this pattern in the later stages, following a
power-law behavior much like the one predicted by the Ogilvie
approach.  Finally also the Ogilvie prediction breaks down far away
from the switch-off point of the forcing.

\acknowledgements
We acknowledge the NORDITA dynamo program of 2011 for providing a
stimulating scientific atmosphere.
JES acknowledges the financial support from the Finnish Cultural Foundation.
The computations have been carried out on the
Parallel Computers at the Royal Institute of Technology in Sweden, and 
the facilities hosted by the CSC  -- IT Center for Science in Espoo, Finland, 
who are financed by the  Finnish ministry of education.
This work was supported in part by the the European Research Council under
the AstroDyn Research Project 227952, and the Academy of Finland grants 136189,
140970 (PJK) and 218159, 141017 (MJM).


\end{document}